\pdfoutput=1
\newcommand{\bb}[1]{\ensuremath{\mathbb{#1}}}

\newcommand{\TT}{\ensuremath{\bb T}}
\documentclass[aps,prl,showpacs,twocolumn,10pt,floatfix]{revtex4-1}

\usepackage[pdftex]{graphicx}

\usepackage{amsmath}   % Or whichever packages you want
\usepackage{amssymb}
\usepackage{verbatim}
\usepackage[usenames,dvipsnames]{color}
\usepackage[colorlinks=true,linkcolor=blue]{hyperref}
\def\omitt#1{} %for source-file notes  --TW 18Jun
\usepackage{Tomspeak}
\def\bodybasis{$\hat e_1, \hat e_2, \hat e_3$ }

\begin{document}
\author{Brian Moths}
\email{bmoths@uchicago.edu}
\author{T. A. Witten}
\affiliation{Department of Physics and James Franck Institute,
University of Chicago, Chicago, Illinois 60637, USA}
\pacs{05.45.-a, 82.70.Dd, 87.50.ch}

%%%%%%%%%%
\title{Full alignment of colloidal objects by programmed forcing}
%%%%%%%%%%%
\date{\today}

\begin{abstract} 
{By analysis and simulation we demonstrate two methods for achieving complete orientational alignment of a set of identical, asymmetric colloidal objects dispersed randomly in a fluid. Sedimentation or electrophoresis in a constant field can lead to partial alignment, in which the objects rotate about a common body axis, but the phases of rotation for these objects are random.  We show that this phase disorder can be removed by two forms of programmed forcing.  First, simply alternating the forcing between two directions reduces the statistical entropy of the orientation arbitrarily.  Second, addition of a small rotating component to the applied field in analogy to magnetic resonance can lead to phase locking of the objects' orientation.  We identify conditions for alignment of a broad class of generic objects and discuss practical limitations.
}
\end{abstract} 

\maketitle

Whenever an object responds to external forcing by periodic motion, this response gives 
potential utility in characterizing the object, manipulating 
it, or using it as a probe.  A prime example is the response 
of a nuclear spin to a static magnetic field.  The response to 
the field enables one to characterize the molecular 
environment of the spin\cite{Seco:2004fk} and to control the quantum 
state of the molecules\cite{Happer:1984uq}.  Analogs using 
electrical or mechanical oscillations are well 
known\cite{Vion:2002ys, Fuchs:2012kx, Abbott:2004vn}.  A further feature 
of magnetic resonance is that the response of many identical 
spins can be manipulated to be {\it coherent}: the spin states 
of all the atoms are identical  so that they all oscillate in 
concert\cite{Purcell_Pound}.  Coherence brings further 
benefits: it enhances the observed signal and it provides 
further information about the constituent objects\cite{Riek:1999zr}.  
Moreover, once coherence is achieved, the objects can respond 
as one to further external forcing\cite{Bardeen:1997ly}.

			These well-known examples rely on a 
sharply resonant response by the individual objects.  Here we examine analogous phenomena in the complementary domain of purely dissipative 
responses of identical, asymmetric colloidal objects dispersed 
in a fluid.  In the simple case of sedimentation, the object 
is pulled by gravity acting at the center of buoyancy and by 
hydrodynamic drag forces over its surface.  
For asymmetric objects the drag forces generally produce 
rotation as well as translation.  This rotational 
sedimentation effect has aroused recent interest as a means of  
organizing colloidal objects\cite{Krapf09, Gonzalez04, 
Makino03,Makino05,Andreev10}.   For many objects, the 
sedimenting force leads to uniform rotation about a specific 
direction in the object which aligns with the force.  
Conditions for this ``axial alignment" have been identified 
\cite{Gonzalez04} and related to object 
shapes\cite{Makino05,Krapf09}.  Any external field producing 
motion in the fluid, such as electrophoresis, produces 
analogous rotational effects\cite{AjdariLong}, as described 
below. 

			This axial alignment under constant 
forcing is necessarily incomplete.  Even when all the objects 
are rotating at the same rate about the same body axis, they 
all have arbitrary angular orientations about the aligning 
axis.  Here we show how programmed, time-dependent forcing can remove this disorder so that all the objects are rotating together coherently.  

			We explain these effects in the simple context of sedimentation of rigid, asymmetric, noninteracting colloidal objects.  We first state the equation of motion that governs rotation of each object.  Then we demonstrate alignment in the simplest case where the force simply alternates between two different directions.  Under mild conditions this alternating forcing leads to continually improving alignment.  We quantify this alignment in terms of statistical entropy, showing that on average it decreases indefinitely.  We next demonstrate alignment via a rotating transverse force.  Finally we discuss generalizations,  experimental implementations and practical limitations of this method.
	
     For simplicity we consider a set of identical objects subjected to a sedimentation force $\vec{F}(t)$ acting at a ``forcing point" $P$ in the object.  We consider the regime of creeping flow \cite{Happel65}, in which inertial forces are negligible and the force transmitted to a moving object by the medium is proportional to the object's velocity. A rotating rigid body with center of mass velocity $\vec{v}$ and angular velocity $\vec{\omega}$ experiences a proportional hydrodynamic force $\vec{F}$ and torque $\vec{\tau}$. In terms of the hydrodynamic radius $R$ of the object, the proportionality can be written in terms of a dimensionless block matrix:
\begin{equation} 
\label{eq:ofmotion}
\left [ \begin{array}{cc}\vec{v}\\
 \vec \omega R\end{array} \right ]= \frac{1}{6 \pi \eta R}\left ( \begin{array}{cc}
	\mathbb{A} & \TT^{T} \\
	\TT & \mathbb{S} \end{array} \right ) \left [ \begin{array}{cc}\vec{F}\\
 \vec \tau/R 
 \end{array} \right ]
\end{equation}
where $\eta$ is the viscosity of the fluid.  For simplicity below we will use units such that $6\pi\eta$ and $R$ are unity.
By choosing the forcing point $P$ as our origin, we remove any external torque, so that any rotation arises entirely from the $3\times 3$ ``twist matrix" $\TT$: $\vec{\omega}=\TT\vec{F}.$ The change of $\TT$ owing to this $\vec{\omega}$ can be written\cite{Krapf09}:
\begin{equation}
\label{eq:TmatrixMotion}
\dot{\bb{T}} = [\vec\omega^\times, \bb{T}],
\end{equation}
Here we use the notation $\vec{\omega^\times}$ to denote the antisymmetric tensor corresponding to the vector $\vec{\omega}$: $[\vec{\omega}^\times]_{ij}\definedas-\epsilon_{ijk}~\omega_k$. Alternatively, in  a rotating frame that is fixed in the object, $\TT$ becomes constant and $F$ rotates: $\dot{F}=-\omega\times F$. 

\begin{figure}
\hskip -2mm

\hbox{
	\vbox{\hsize=18mm \vskip 5mm \includegraphics[width=17mm]{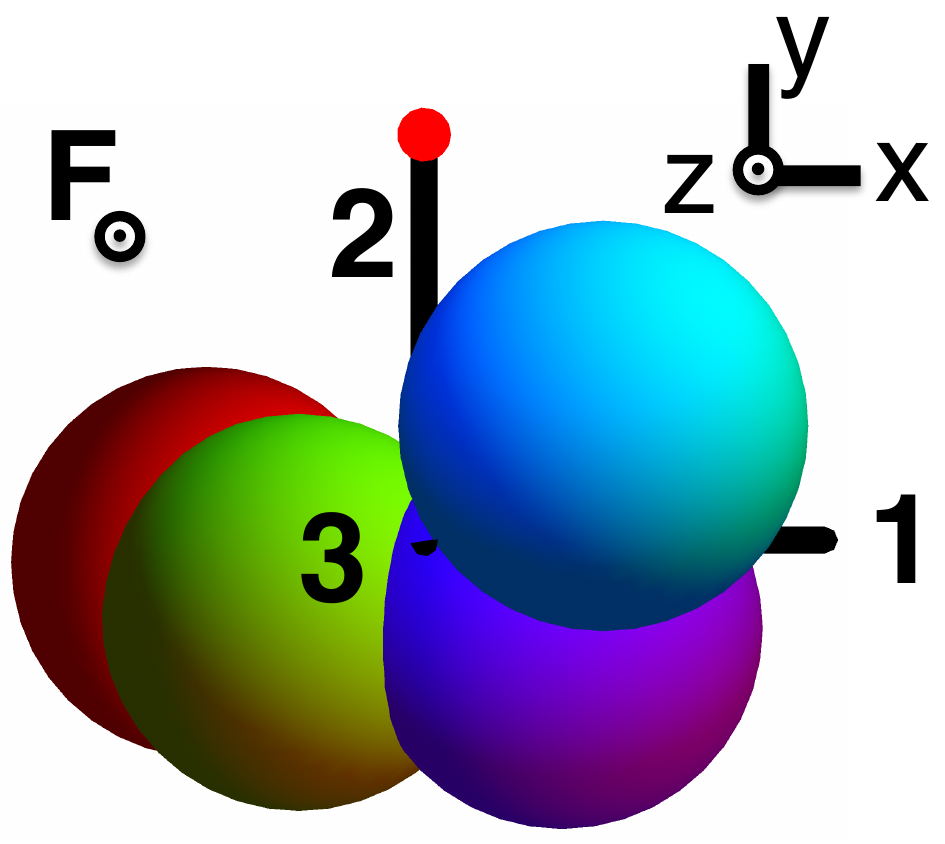}\vskip 3mm}
	\hskip 3mm
	\includegraphics[width=65mm]{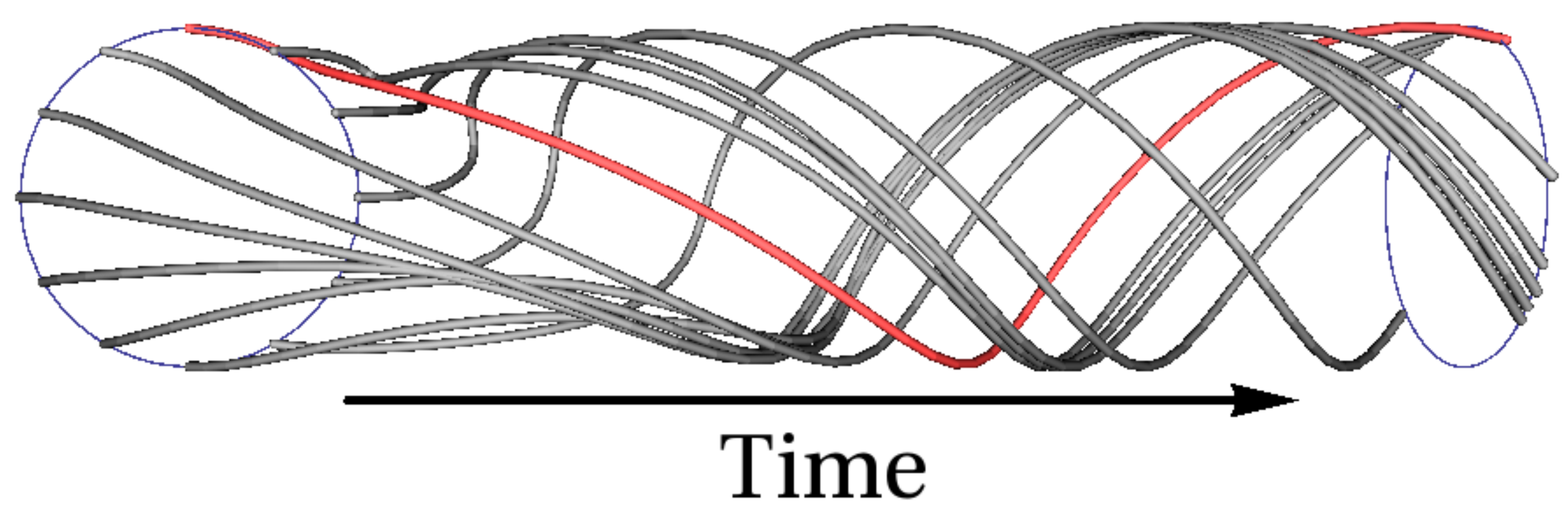}
}

\vskip -22mm 
\hbox to \hsize{\hskip 0mm a \hskip 18mm b\hfill } %\vskip -4pt
\vskip 21mm
\includegraphics[width=\hsize]{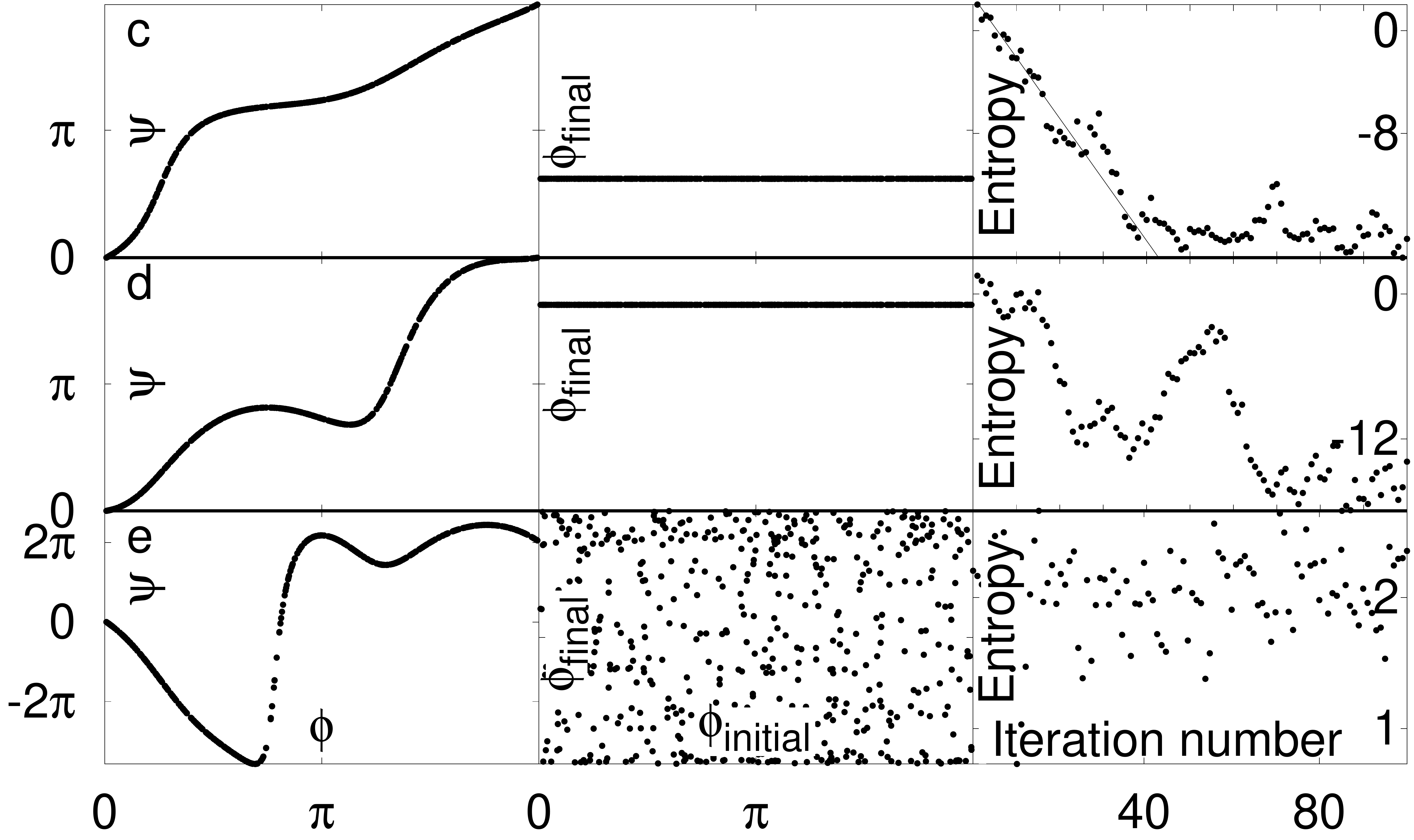}
\caption{ a) The conjoined-sphere object referenced in the main text axially aligned to a downward force $\vec F$, viewed from slightly below the horizontal. A lab basis $x$-$y$-$z$  and a body basis \bodybasis, centered at the forcing point, are shown. The azimuthal angle $\phi$ is 0. The object's $\TT$ matrix in \bodybasis co-ordinates was computed using the HYDROSUB software package \cite{Hydrosub}. 
A colored dot marks the $\hat e_2$ axis.   b) Motion after a switch in force direction. The red line illustrates the trajectory of the colored dot after the forcing direction has switched to the x direction. Eleven other trajectories for objects initially rotated by different $\phi$'s are also shown.  After re-alignment the points have become non-uniformly distributed on the new circle. c), d) e): sample results of repeated rocking by 90 degrees as described in the text.  Left-hand graphs show the mapping function $\psi(\phi)$.  Center graphs show the angles $\phi$ for 100 randomly oriented copies of the object after 100 switches. Right-hand  graphs show the progression of the entropy $H$ with repeated rocking. Final low values are consistent with the noise floor of the numerics .   c) shows the results of a monotonic $\psi$ function. Line indicates the predicted slope from (\ref{eq:dhpsiav}). d) shows the results for the object shown in a, e) shows the results for a strongly non-monotonic $\psi(\phi)$ whose entropy does not decrease.}
\label{fig:rocking}
\end{figure}

In what follows we restrict our attention to a subclass of $\TT$'s---denoted ``axially aligning"---that align in a unique direction, independent of initial orientation, under constant forcing.  Such $\TT$'s have sufficiently large antisymmetric parts that they have only one real eigenvalue $\lambda_3$ and eigenvector $\vec{v}_3$.  Then $\TT$ aligns with its $\vec{v}_3$ along $\vec{F}$ and rotates at angular velocity $\lambda_3\vec{F}$ \footnote{In practice many non-symmetric objects are axially aligning\cite{Krapf09}. }. An example of one such object, made by four conjoined spheres, is shown in Fig. \ref{fig:rocking}a \footnote{The buoyant masses from closest to farthest are in the ratio 1.34, 0.80, 0.72, 1.14. For this object, up to a numerical factor, 
$$
\TT \aboutequal
\left(
\begin{array}{ccc}
 -0.46 & -0.71 & 0. \\
 0.25 & -0.41 & 0. \\
 0.23 & 0. & 1. \\
\end{array}
\right)$$
In a $10^4$ g centrifuge this object with spheres of 1 micron diameter and specific gravity of 1.1 in water would rotate at 25. rad/sec.
}.

We first consider a minimal forcing program: a mere switch of the forcing direction by a ``rocking angle" $\theta$ from its initial direction along the $z$ axis. Fig. \ref{fig:rocking}b illustrates the result of this switch.  An initially uniform distribution of phase angles $\phi$ becomes nonuniform; the transient relaxation has reduced the disorder.

More explicitly, we may define a basis of unit vectors \bodybasis   in the object  whose $\hat e_3$ axis is the direction that aligns with the  force $\vec F$, initially along the $z$ axis in the lab frame (Fig. \ref{fig:rocking}a). At a given moment this object's $\hat e_2$ vector makes an azimuthal angle $\phi$ with the lab's $y$ axis.  This $\phi$ increases steadily in time as explained above.  The $\phi$'s of the different objects are presumed to be uniformly distributed.  After $\vec F$ has switched into the $x$-$z$ plane and the objects have re-aligned, their $\hat e_2$ vectors again differ only in their azimuthal angle with the $y$ axis, which we denote $\psi$.  Evidently an object's angle $\psi$ at a time $t$ after the switch depends on the angle $\phi$ immediately before the switch: $\psi = \psi_t(\phi)$.  As the figure illustrates, the $\psi$ angles are no longer uniformly distributed: some $\psi$'s have bunched closer together; others have spread apart.  For convenience we shall choose a time $t$ so that $\psi(0)=0$.  Evidently if the rocking angle $\theta = 0$, the motion is a continuation of the uniform rotation without any switch: $\psi(\phi) = \phi$.  If $\theta$ increases from zero by a small amount,  $\psi(\phi)$ remains close to $\phi$ and $\psi(\phi)$ remains monotonic \cite{bigpaper}. Further, for general $\theta$, as $\phi$ advances by $2\pi$, $\psi$ must advance by the same net amount.  Thus $2\pi = \oint d\phi~ (d\psi/d\phi)$ \cite{bigpaper}.  Typical $\psi(\phi)$'s are shown in Fig. \ref{fig:rocking}c-e.

We may quantify the bunching effect of $\psi$ using the probability distribution function $p(\phi)$ measured after the objects have aligned.  The net effect on the distribution can be quantified using the statistical entropy $H \definedas - \integral p \ln p$.  This $H$ is maximal for uniform probabilities and is small when the probability is concentrated into small regions\cite{Shannon_1948}.  After one switching process, each angle $\phi$ evolves into some $\tilde \phi$.  The passage from $\phi$ to $\tilde \phi$ involves two steps: we first wait a random fraction of the rotational period, so that $\phi$ undergoes a shift by a random angle $\alpha$.  We then switch the force $F$ and allow the objects to realign. The resulting angle $\tilde \phi = \psi(\phi + \alpha)$.  The corresponding probability distribution $\tilde p(\tilde \phi)$ is found using $ \tilde p(\tilde \phi)~ d\tilde \phi = p(\phi)~ d\phi$, so that $\tilde p(\tilde \phi) = p(\phi)/\psi'(\phi + \alpha)$. The change of $H$ resulting from this process can readily be computed provided $\psi'(\phi)$ is positive for all $\phi$\cite{bigpaper}.  The change $\Delta H$ depends on the shift angle $\alpha$:
\begin{equation}
	\label{eq:dhpsi}
	\Delta H_\alpha = \oint p(\phi) \ln \left( \psi'(\phi+ \alpha)\right) \, d \phi.
\end{equation}

We now consider the net change of $H$ after a sequence of many switches of $F$.  Then the $\alpha$-averaged change of $H$, $\expectation{\Delta H}$ can be written 
\begin{equation}
	\label{eq:dhpsiav}
	\expectation{\Delta H} = \oint 
	\expectation{p(\phi + \alpha)}~~ \ln \left( \psi ' (\phi)\right) \, d \phi.
\end{equation}

The $\expectation{p(\phi + \alpha)}$ is an unimportant positive constant. The remaining integral is a constant that is necessarily negative\cite{Shannon_1948}, owing to the convexity of the logarithm.  Thus after many switches, the entropy must decrease indefinitely on average and the probability $p$ becomes confined to arbitrarily small regions of $\phi$.  As Fig. \ref{fig:rocking}c and d illustrate, the initial set of $\phi$'s often evolve into a single final $\phi$ \footnote{Kaijser \cite{Kaijser1993} argues that the probabilities {\em must} collapse to a single point.}.  Further, this reduction of $H$ often occurs even when $\psi'$ is not positive definite (Fig. \ref{fig:rocking}d).  Given a mixture of two or more alignable species, this method aligns each species.  However, one readily finds examples $\psi(\phi)$ where the entropy does not decrease (Fig. \ref{fig:rocking}e).  In any case the direction of alignment is not controlled in this method.

A second method of forcing can achieve alignment in a controlled direction.  We add a rotating transverse force to the original static force, so that the force vector is tilted at an angle $\theta$ from the $z$ axis.  Thus $\vec F$ rotates at a constant angular velocity $\vec \Omega = |\Omega|~\hat z$.  
\begin{equation}
	\label{eq:fxy}
	\vec{F}(t)=|F|\left(~\hat{z} \cos\theta ~+~  [\hat{x} \cos\Omega t + \hat{y} \sin\Omega t]\sin\theta~\right).
\end{equation}

In a {\em co-rotating} frame rotating at angular velocity $\vec \Omega$, the force $\vec F$ becomes constant.
With a proper choice of $\Omega$ and $\theta$ the objects too may evolve into a state of co-rotation with the force, with a common orientation. Co-rotation requires that $\TT$ remain in an orientation such that $\vec \Omega = \TT \vec F$. As noted above, a fixed force with $\theta = 0$ leads to co-rotation with $\vec \Omega = \lambda_3 \vec F$. In the \bodybasis body reference frame, denoted with subscript $b$, co-rotation means $\vec \Omega_b = \TT_b \vec F_b$ for some constant $\vec F_b$ and corresponding $\vec \Omega_b$.    The forcing parameters required are evidently a) $|F|^2 = |F_b|^2$, b) $|\Omega|^2 = |\Omega_b|^2$ and c) $\cos \theta = \hat F_b \cdot \hat \Omega_b$.  

\begin{figure}
\hbox to \hsize{\vbox{\hsize=30mm \includegraphics[width=30mm]{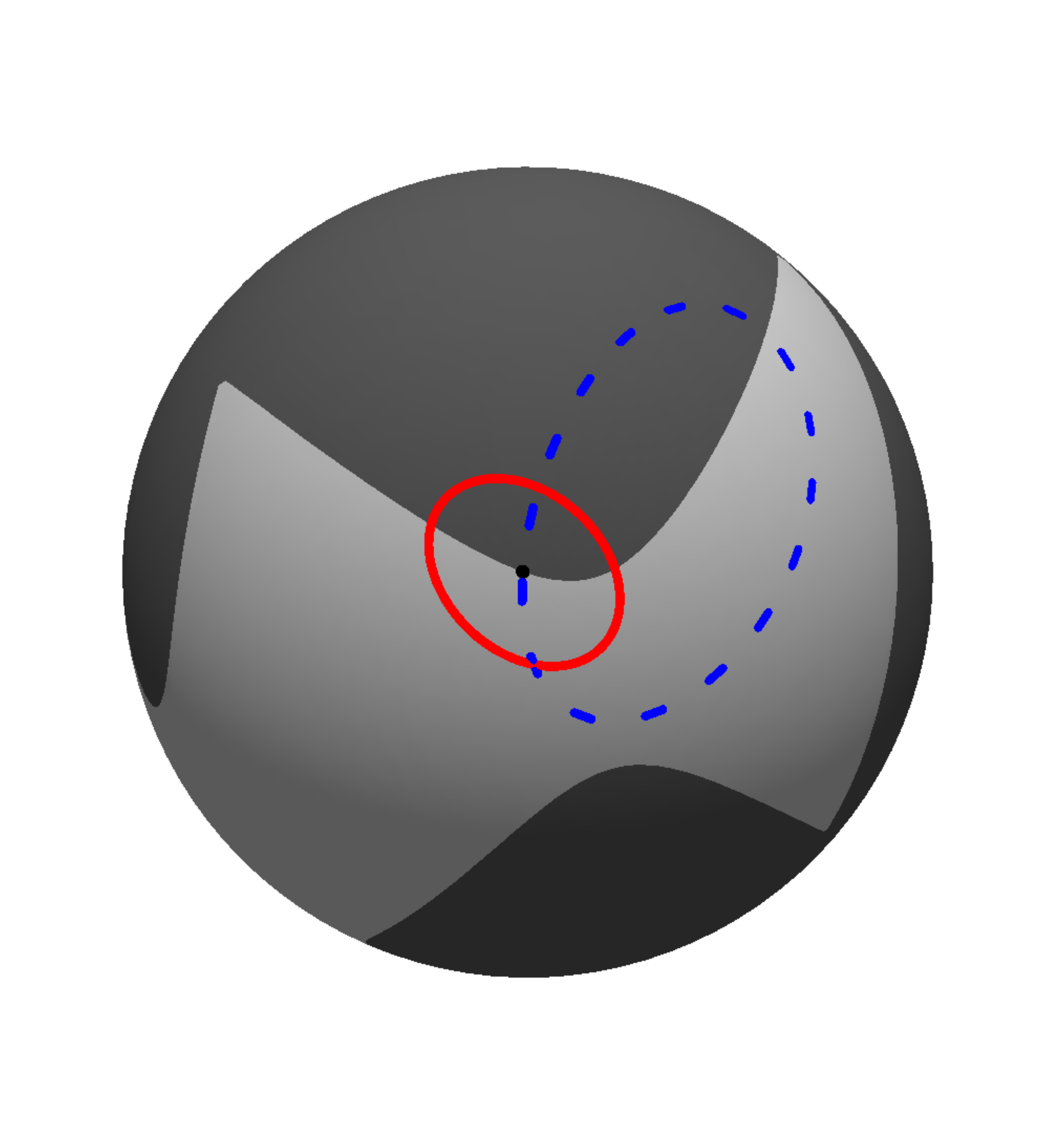}\break
\includegraphics[width=30mm]{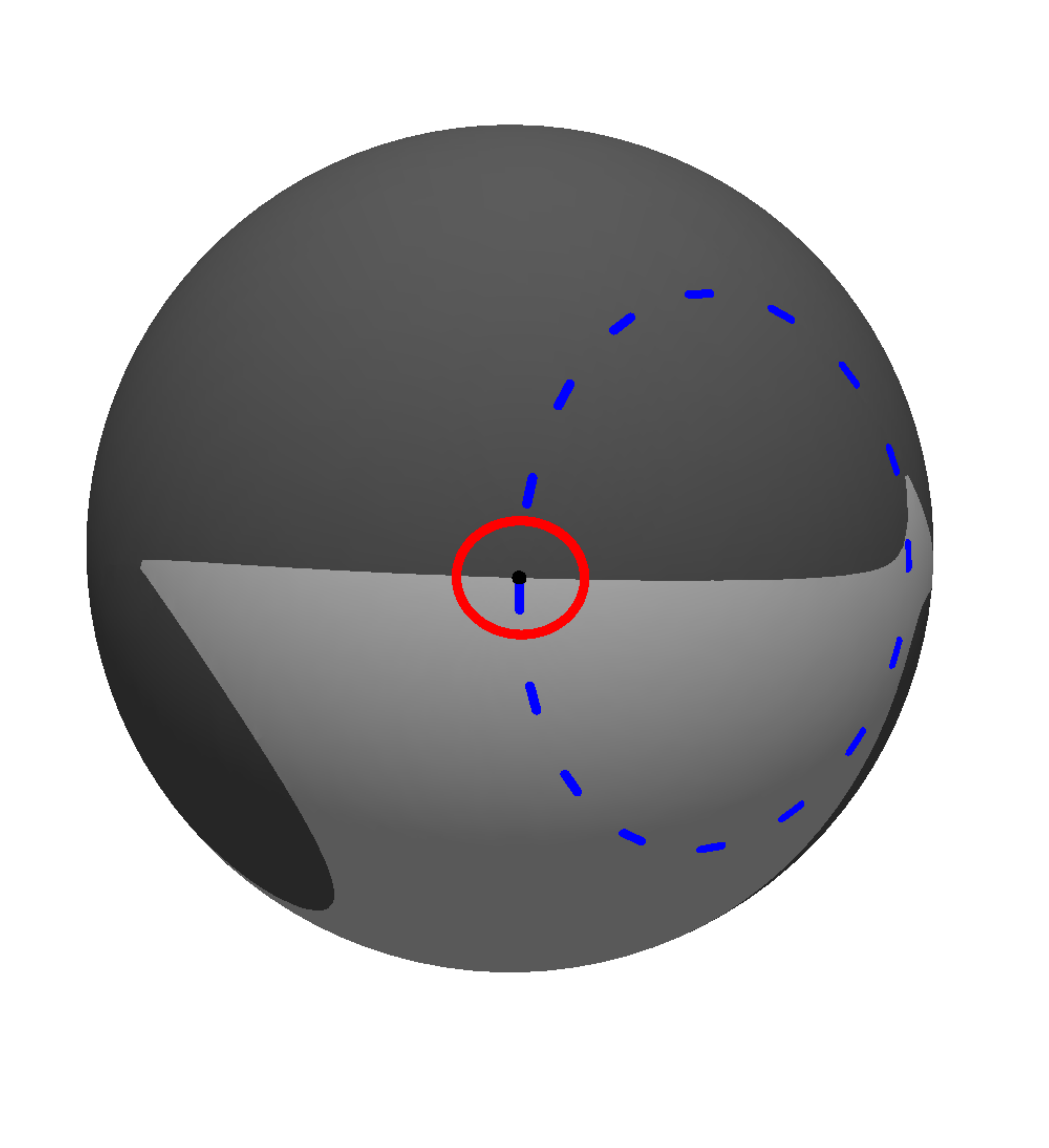}\break 
\includegraphics[width=30mm]{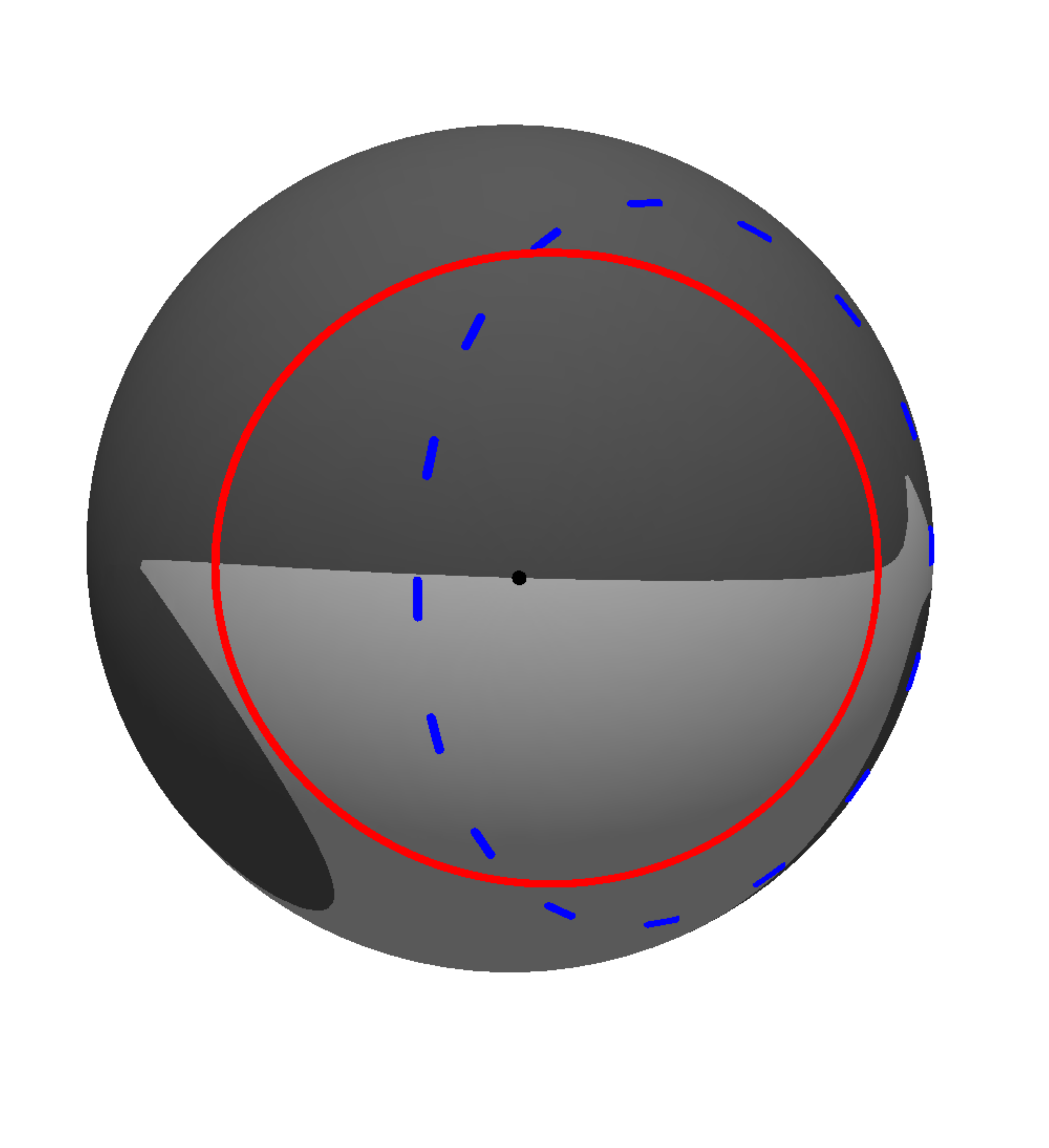}\vskip 0mm}
\includegraphics[height=90mm]{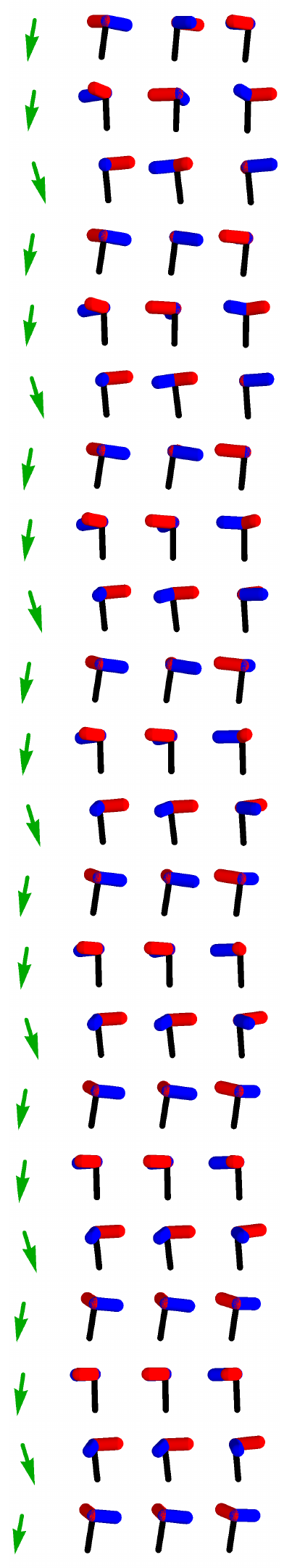} 
\hskip 1pt\includegraphics[height=90mm]{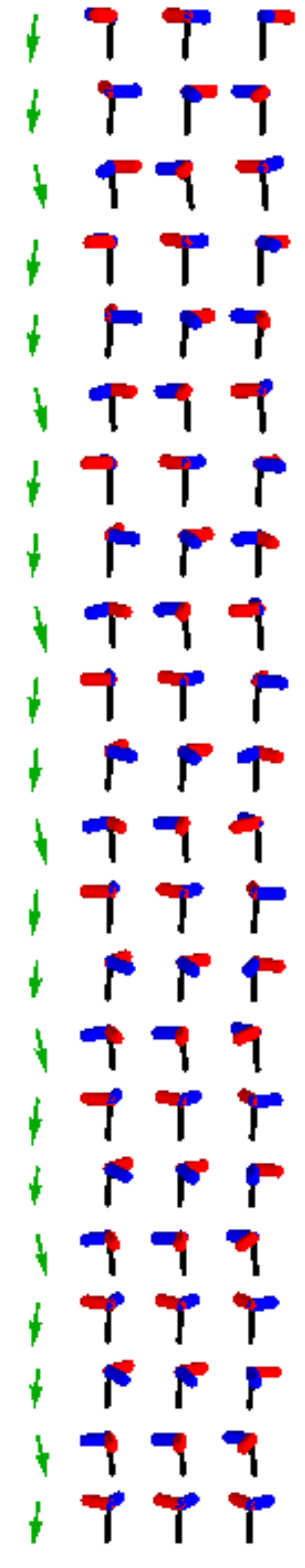}
\hskip -1pt\includegraphics[height=90mm]{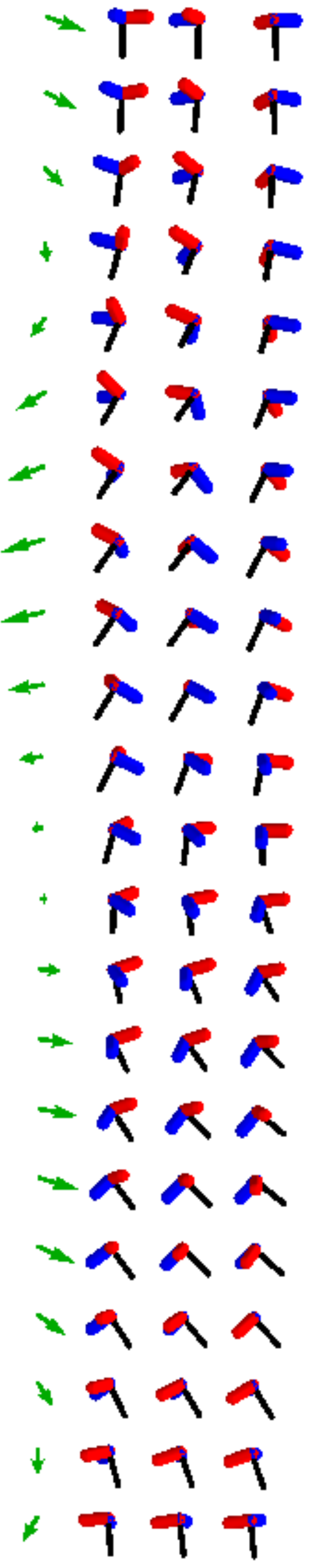}
}
\vskip -3.65in
\hskip .45in {\bf d \hskip 17mm e \hskip 16mm f}
\vskip -.1in
\hskip -3.2in \vbox{ \bf a \vskip 1.1in b \vskip 1in c \vskip 1.1in}

\caption{a) The sphere of possible directions of the body frame force $\vec{F_b}$ for the object of  Fig. \ref{fig:rocking}a.  The solid curve gives the $\hat F_b'$\,s satisfying constraint c in the text, enforcing an angle $\theta = 0.3$.  The dashed curve gives the $\hat F$'s satisfying constraint b in the text, enforcing $|\Omega| = \lambda_3 |F|$. The intersections of the two curves mark co-rotating states.  The light shading marks regions of stable fixed points $\hat F_b$ for this object. b), Same, for the $\TT$ used for Fig. \ref{fig:rocking} e with $\theta=0.2$.  c) Same as b, with $\hat F_b$ chosen to lie far from the aligning direction $\hat e_3$ to illustrate alignment with large tilt angle.  Here $|\Omega \notequal \lambda_3 |F|$; instead, $\vec \Omega_b = \TT_b \vec F_b$.  d) Time sequence of orientations using the object and alignment protocol described in a. Three arbitrarily oriented copies of the object, represented by their \bodybasis bases are shown. Colored arrows indicate the direction of forcing.  The three objects evolved to a common final orientation. e) Similar time sequence for the $\TT$ matrix used for Fig. \ref{fig:rocking} e. f) Similar time sequence for the $\TT$ matrix and forcing condition shown in c.
}
\label{fig:sphere1}
\end{figure}

With this same choice of parameters other co-rotating $\vec F$'s are also possible.  We denote these as $\vec F_b'$.  Because of condition a)  such $\vec F_b'$'s have the same magnitude as $\vec F$, so that the only difference in the $\vec F_b$'s is in their direction $\hat F_b'$.  In terms of the matrix $\TT_b$, condition b) can be written 
$\hat F_b' ~\TT_b{}^T \TT_b~ \hat F_b' =\hat F_b ~\TT_b{}^T \TT_b~ \hat F_b ~(= |\Omega|^2/|F|^2)$. (If $\hat F_b'$ satisfies this condition, so does $-\hat F_b'$.) This condition restricts $\hat F_b'$ to two (closed) curves on the unit sphere like the dashed line in Fig. \ref{fig:sphere1}a-c.  Likewise condition c) reads
$ \hat F_b' ~\TT_b~  \hat F_b'~/|\TT_b \hat F_b'| = \hat F_b ~\TT_b~  \hat F_b~/|\TT_b \hat F_b| ~(= \cos \theta) $.  This condition restricts $\hat F_b'$ to a second pair of curves on the unit sphere.  Any intersection of these curves represents an $\hat F_b'$ that co-rotates with the given forcing. 

These compatible $F_b$'s can readily be found when the tilt angle $\theta$ is small. Figs. \ref{fig:sphere1}a-c show the behavior of the co-rotating $\hat F_b'$\,s as one increases $\theta$ from 0 with $|\Omega| = \lambda_3 F$. The condition-b) curves, enforcing the magnitude of $\Omega$, are thus independent of $\theta$  and do not change.  These curves pass through $\hat e_3$ and $-\hat e_3$, \ie\ the $\hat F_b$ for $\theta = 0$. Condition c), enforcing $\cos \theta$ requires $\hat F_b \parallel \hat e_3$ when $\theta = 0$.  As $\theta$ increases, the condition-c) curves expand to small rings encircling $\plusorminus \hat e_3$. Each ring must intersect its condition-b) curve twice.  The four co-rotating $\hat F_b'$\,s are then two adjacent pairs near $\plusorminus \hat e_3$.   In what follows we consider only the two intersections adjacent to the stable $+\hat e_3$ direction

To achieve full alignment, an arbitrary initial state must evolve into one of these two co-rotating states.  We argue that this alignment occurs generally for sufficiently small $\theta$.  We consider the motion of $\TT$ in the co-rotating frame, in which $\vec F$ and $\vec \Omega$ are fixed.  We first align $\TT$ axially using a constant $\vec F$. We can express the orientation of $\TT$ using a rotation vector $\vec \eta$.  We distinguish angular rotations $\eta_3$ along the $\vec \Omega$ axis from rotations $\eta_\perp$ perpendicular to it. In this initial state, $\eta_\perp = 0$, while the orientation $\eta_3$ about the aligning axis  is arbitrary:  axial alignment has restricted $\vec \eta$ to a closed one-dimensional curve of possible values.  A small rotation $\eta_3$ has no further effect on the motion, but a small rotation $\eta_\perp$ leads to a stable return to $\eta_\perp = 0$.

Now we increase $\theta$ to a small nonzero value. Then $\vec \eta$ is no longer constant.  The time derivative $\dot{\vec \eta}_\theta(\vec \eta)$ differs from $\dot{\vec \eta}_0(\vec \eta)$ by a small, smooth perturbation. We now suppose that $\eta_\perp$ converges to some $\eta_3$-dependent value near 0.  The remaining motion, is along a one-dimensional closed curve, slightly distorted from the neutral curve found at $\theta= 0$.  We may express this residual one-dimensional motion using the co-ordinate $\eta_3$, whose time derivative is some function ${\dot \eta}_{3 \; \! \theta}(\eta_3)$. The two fixed points identified above are necessarily fixed points $\eta_3^*$, with $\dot \eta_3(\eta_3^*)= 0$.  Since ${\dot \eta}_{3 \; \! \theta}$ is a smooth function on the curve, its derivatives at these two fixed points are necessarily opposite in sign; hence opposite in stability.  Thus in this picture all $\eta_3$ must must converge to the stable $\eta_3^*$.  Numerical studies like those of Fig. \ref{fig:rocking} confirmed this finding for numerous asymmetric $\TT$'s . 

A local stability analysis confirms the opposite stabilities of the two fixed points $\hat F_b$ near $\hat e_3$.  Starting from a given forcing with a corresponding $\hat F_b$, we rotate $\TT$ by a slight angular displacement $\vec \eta$ from its fixed-point state $\TT^*$, in the co-rotating frame so that $\TT = \TT^* + [\vec \eta^\times, \TT^*]$. The undisplaced state $\vec \eta = 0$ is co-rotating, so that $\dot{\vec \eta} = 0$.  Near $\vec \eta = 0$, $\dot{\vec \eta}$ varies linearly with $\eta$: 
$\dot{\vec \eta} = \bb K \vector \eta$
for some ``stability matrix" $\bb K = [\bb{T}^* \vec{F}^\times-(\TT^* \vec F)^\times]$ \cite{bigpaper}.  From this $\bb K$ one may determine by standard methods \cite{strogatz2001nonlinear} whether the motion returns stably to the co-rotating state at $\TT^*$ after the small displacement $\vec \eta$; and, indeed, one finds that the two fixed points of the previous paragraph have opposite stabilities. Figs. \ref{fig:sphere1}a-c use this $\bb K$ to determine which $\hat F_b$ represent stable co-rotating states. The figure shows large stable and unstable regions of $\hat F_b$. We note that the aligning axis $\hat e_3$, being neutrally stable, lies at the boundary between the stable and unstable regions.

These alignment methods are applicable for any situation where an object responds to a vector field by rotating.  For example, any asymmetric object with an electrophoretic mobility has such a rotational response\cite{AjdariLong}.  Any proportional response is necessarily governed by a $\TT$ matrix like that above.  This is true for deformable objects whenever the driving force is weak enough to avoid deformation.  It is true for fluctuating shapes whenever the driving is weak enough to average over the fluctuations of $\TT$. 

Complete alignment brings benefits that are not possible in  axial alignment.  Once a set of objects have been completely aligned, they must respond identically to subsequent forcing.  Then \eg by  gradually increasing the $\Omega$ of Eq. \ref{eq:fxy} one may achieve phase locking without knowing about the $\TT$ of the objects in advance \cite{bigpaper}. Further extensions of programmed phoresis hold promise for orienting non-axially-aligning objects not considered here.

In practice, alignment is degraded by rotational diffusion, characterized by a relaxation time $\tau_D$, which scales as the cube of the object's size $R$ \cite{Happel65}. The driving forces must be sufficiently strong to produce rotational speeds $\omega$ much larger than $1/\tau_D$.  Thus alignment is strongly degraded as $R$ is decreased.  The object of Fig. \ref{fig:rocking}a has $\Omega \tau_D \aboutequal 260$.  Under electrophoresis an object with a nominal mobility of $10^{-8}$ meters/(volt sec) and field of $10^4$ volts/m, would need to be over 10 microns in size to give comparable $\Omega \tau_D$.  Thus these methods require objects of near-micron scale or larger.

Unless the objects are very dilute, their mutual hydrodynamic interactions would be significant: an object B is advected by the perturbed Oseen flow around a nearby object A \cite{chaikinFLuidized}.  Only the gradient of this flow {\em rotates} object B. The rotational perturbations on an object are thus shorter range and hence weaker than are translational hydrodynamic interactions.  Ordinary electrophoresis produces no Oseen  flow in the host fluid\cite{AjdariLong}; thus electrophoresis is less subject to interaction effects.

The authors thank Michael Solomon for his suggestion that this work is applicable to electrophoresis,  Nathan Krapf for valuable guidance and Michael Cates for useful comments on the manuscript. This work was supported in part by the National Science Foundation's MRSEC Program under Award No. DMR-0820054.

\bibliography{savedrecs}
\end{document}